\documentclass[preprintnumbers,article,amsmath,amssymb,floatfix,10pt,prd,twocolumn,
superscriptaddress,nofootinbib]{revtex4-2}
\usepackage{bm}
\usepackage{amsfonts}
\usepackage{latexsym}
\usepackage[latin1]{inputenc}
\usepackage{graphicx}
\usepackage{amsmath}
\usepackage{palatino}
\usepackage{mathpazo}
\usepackage{textcomp}
\usepackage{float}
\usepackage{booktabs}
\usepackage{dcolumn}
\usepackage{ragged2e}
\usepackage{hyperref}
\hypersetup{colorlinks,citecolor=blue}
\hypersetup{colorlinks=true,linkcolor=red,filecolor=magenta,    urlcolor=blue}
\usepackage{amsmath}
\usepackage{xcolor}
\usepackage{orcidlink}
\usepackage{epsfig}
\usepackage[caption=false]{subfig}
\usepackage{commath}
\usepackage{stackengine}
\newcommand\ocirc[1]{\ensurestackMath{\stackon[1pt]{#1}{\mkern2mu\circ}}}
\def\jnl@style{\it}
\def\aaref@jnl#1{{\jnl@style#1}}

\def\aaref@jnl#1{{\jnl@style#1}}

\def\aj{\aaref@jnl{AJ}}                   
\def\apj{\aaref@jnl{ApJ}}                 
\def\apjl{\aaref@jnl{ApJ}}                
\def\apjs{\aaref@jnl{ApJS}}               
\def\apss{\aaref@jnl{Ap\&SS}}             
\def\aap{\aaref@jnl{A\&A}}                
\def\aapr{\aaref@jnl{A\&A~Rev.}}          
\def\aaps{\aaref@jnl{A\&AS}}              
\def\mnras{\aaref@jnl{Mon.~Not.~Roy.~Astron.~Soc.}}             
\def\prd{\aaref@jnl{Phys.~Rev.~D}}        
\def\prc{\aaref@jnl{Phys.~Rev.~C}}  
\def\prl{\aaref@jnl{Phys.~Rev.~Lett.}}    
\def\qjras{\aaref@jnl{QJRAS}}             
\def\skytel{\aaref@jnl{S\&T}}             
\def\ssr{\aaref@jnl{Space~Sci.~Rev.}}     
\def\zap{\aaref@jnl{ZAp}}                 
\def\nat{\aaref@jnl{Nature}}              
\def\aplett{\aaref@jnl{Astrophys.~Lett.}} 
\def\apspr{\aaref@jnl{Astrophys.~Space~Phys.~Res.}} 
\def\physrep{\aaref@jnl{Phys.~Rep.}}      
\def\physscr{\aaref@jnl{Phys.~Scr}}       
\def\commat{\aaref@jnl{Comm.~Math.~Phys.}}              
\def\science{\aaref@jnl{Science}}               
\def\cqg{\aaref@jnl{Classical Quant.~Grav.}}            
\def\jpcs{\aaref@jnl{JPCS}}                                     
\def\ijmpd{\aaref@jnl{Int.~J.~Mod.~Phys.~D}}                    
\def\grg{\aaref@jnl{Gen.~Relat.~Gravit.}}               
\def\rpp{\aaref@jnl{Rep.~Prog.~Phys.}}          
\def\npa{\aaref@jnl{Nucl.~Phys.~A}}        
\def\lrr{\aaref@jnl{Living Rev.~Rel.}}                   
\def\jcap{\aaref@jnl{J.~Cosmology Astropart.~Phys.}}    
\def\rmp{\aaref@jnl{Rev.~Mod.~Phys.}}   
\def\epjc{\aaref@jnl{Eur.~Phys.~J.~C}}
\def\plb{\aaref@jnl{~Phy.~Lett.~B}}
\def\mpla{\aaref@jnl{Mod.~Phy.~Lett.~A}}
\def\arxiv{\aaref@jnl{arxiv.org}}

\allowdisplaybreaks[1]

\begin{document}
\color{black}       
\title{Reply to ``Comment on `Energy Conditions in $f(Q)$ gravity''}
\author{Sanjay Mandal\orcidlink{0000-0003-2570-2335}}
\email{sanjaymandal960@gmail.com}
\affiliation{Department of Mathematics, Birla Institute of Technology and
Science-Pilani,\\ Hyderabad Campus, Hyderabad-500078, India.}
\author{P.K. Sahoo\orcidlink{0000-0003-2130-8832}}
\email{pksahoo@hyderabad.bits-pilani.ac.in}
\affiliation{Department of Mathematics, Birla Institute of Technology and
Science-Pilani,\\ Hyderabad Campus, Hyderabad-500078, India.}
\author{J.R.L. Santos\orcidlink{0000-0002-9688-938X}}
\email{joaorafael@df.ufcg.edu.br}
\affiliation{UFCG - Universidade Federal de Campina Grande - Unidade Acad\^{e}mica de F\'isica,  58429-900 Campina Grande, PB, Brazil.}
\date{\today}
\begin{abstract}
Recently, Avik De and L.T. How \cite{avik} claim that we have not followed an effective theory approach for  $f(Q)$ gravity, missing  relevant terms to compute the energy conditions \cite{sanjay}. They also state that we did not present how the pressure and energy density follow a set of energy condition criteria. In this reply we show how the effective density and effective pressure found in \cite{avik} yield to the same Fridmann Equations presented in our article. Therefore, we claim that the comment \cite{avik} only introduces an equivalent path to derive the Friedmann Equations and that the extra terms found by Avik De and L.T. How are consequence of their specific definition for their effective density and effective pressure. However, such terms do not change the equations for $\rho$ and $p$. Consequently, these extra terms are not going to change the energy constraints presented by us in \cite{sanjay}.


\end{abstract}

\maketitle

\date{\today}
The search for a complete theory of gravity has been impelling the proposals of new theories beyond general relativity. Among this effort, we highlight the so-called symmetric teleparallel gravity or $f(Q)$ gravity, where $Q$ is the trace of the nonmetricity tensor. Such a theory was introduced by Jimenez et al. \cite{5}, and investigations on its subjects have been rapidly increased, as well as observational constraints to confront it against standard GR formulation.

One of the proposals to test $f(Q)$ gravity was introduced by us \cite{sanjay}, where we worked on constraints over different energy conditions, establishing the viability of different models. In our approach, the non-trivial features due to the nonmetricity scalar were properly embedded into an effective density and effective pressure.  

 In a recent comment, Avik De and L.T. How \cite{avik} have disputed our approach to derive energy conditions to $f(Q)$ gravity. They claim that we have not followed the effective theory approach to study the energy conditions and also missed some terms in our study. In this reply, we are going to show that in fact the approach presented by Avik De and L.T. How yield to the same set of equations for density and pressure that we found in \cite{sanjay}, corresponding to an equivalent path to derive energy conditions. 
 
Let us start with the field equation, which is incorporated in our study \cite{sanjay}, and whose form is
\begin{multline}
\label{1}
\frac{2}{\sqrt{-g}}\nabla_{\gamma}\left( \sqrt{-g}f_Q {P^{\gamma}}_{\mu\nu}\right)+\frac{1}{2}g_{\mu\nu}f\\
+f_Q\left(P_{\mu\gamma i}{Q_{\nu}}^{\gamma i}-2Q_{\gamma i \mu}{P^{\gamma i}}_{\nu} \right)=-T_{\mu\nu}\,.
\end{multline}

Later, the covariant formulation to study the $f(Q)$ gravity was proposed by Dehao Zhao \cite{zhao} in 2022. There the field equation is reconstructed using covariant formalism, which is given by

\begin{equation}
\label{2}
f_Q \ocirc{G}_{\mu\nu}+\frac{1}{2}g_{\mu\nu}(f_Q Q-f)+2f_{QQ}(\delta_{\lambda}Q)P^{\lambda}_{\mu\nu}=T_{\mu\nu}.
\end{equation}

This formulation is followed by Avik and How \cite{avik}, and it is different from our proposal \cite{sanjay} as well as from many articles in the same subject, such as \cite{3,4,5,6} .


Despite this difference in deriving the field equations for $f(Q)$ gravity, the Friedmann equations computed from them should be the same. It is relevant to inform that the Friedmann equations for $f(Q)$ gravity are a crucial point to derive the energy conditions coming from the  Raychaudhury equations. 

In this answer, we are going to show that the field equations
\begin{equation} \label{eq01}
\ocirc{G}_{\mu\nu}=\frac{1}{-f_Q}T^{eff}_{\mu\nu},
\end{equation}
with
\begin{equation}
\ocirc{G}_{\mu\nu}=-(2\dot{H}+3\,H^{2})\,g_{\mu\nu}-2\dot{H}u_{\mu}u_{\nu}\,,
\end{equation}
and
\begin{equation} \label{eq02}
G_{\mu\nu}=T^{eff}_{\mu\nu},
\end{equation}
yield to the same set of Friedmann equations. Here we clarify that Eq. \eqref{eq01} was presented in the refereed Comment, whereas we worked with Eq. \eqref{eq02} in  \cite{sanjay}.

So, let us start by taking the energy-momentum tensor presented in the refereed Comment, whose form is
\begin{equation} \label{eq03}
T_{\mu\nu}^{eff}=T_{\mu\nu}+\left(\frac{f}{2}-3\,H^{2}f_{Q}\right)\,g_{\mu\nu}-24H^{2}\,\dot{H}\,f_{QQ}\,\left(g_{\mu\nu}+u_{\mu}u_{\nu}\right)\,,
\end{equation}
where we choose $\kappa =1$ for the sake of simplicity. The last equation can be rewritten in the following form
\begin{equation}
T_{\mu\nu}^{eff}=T_{\mu\nu}+\left(\frac{f}{2}-3\,H^{2}f_{Q}\right)\,g_{\mu\nu}-2\dot{f}_{Q}H\,\left(g_{\mu\nu}+u_{\mu}u_{\nu}\right)\,,
\end{equation}
once
\begin{equation}
Q=6\,H^2\,,
\end{equation}
and then
\begin{equation}
\dot{f}_{Q}=f_{QQ}\,\dot{Q}=12\,f_{QQ}\,\dot{H}\,H\,.
\end{equation}

Taking $00$ and $ii$ components from Eq. \eqref{eq03}, the authors of the refereed Comment found that (see Eqs. (13) and (14) in \cite{avik})
\begin{equation} \label{eq04}
\rho^{eff}=\rho-\frac{f}{2}+3\,H^2\,f_Q\,,
\end{equation}
\begin{multline} \label{eq05}
p^{eff} = p+\frac{f}{2}-3H^2f_Q -24 H^{2}\dot{H}f_{QQ}\\
= p+\frac{f}{2}-3H^2f_Q -2 \dot{f}_{Q} H\,.
\end{multline}
Now, taking $00$ and $ii$ components from Eq. \eqref{eq01} we find that
\begin{equation} \label{eq06}
\rho^{eff}=-3\,f_Q\,H^2\,,
\end{equation}
and
\begin{equation} \label{eq07}
p^{eff}=f_Q\,\left(2\dot{H}+3H^2\right)\,.
\end{equation}

Therefore, by comparing Eqs. \eqref{eq04} and \eqref{eq05} to \eqref{eq06} and \eqref{eq07}, we yield to
\begin{equation} \label{eq08}
\rho = \frac{f}{2}-6\,H^2\,f_{Q}\,,
\end{equation} 
and
\begin{equation} \label{eq09}
p=\left(\dot{H}+3\,H^2+\frac{\dot{f}_Q}{f_{Q}}\,H\right)\, 2\,f_Q-\frac{f}{2}\,,
\end{equation}
respectively. Equations \eqref{eq08} and \eqref{eq09} were also presented by us in \cite{sanjay} (see Eqs. (13) and (14) in that paper). Thus, we and the authors of the refereed comment found the same set of Friedmann Equations, which enables one to define effective density and effective pressure, constraining the energy conditions for $f(Q)$ gravity. The present approach also clarifies the proper criteria to write $\rho$ and $p$.

In addition, the effective energy density and pressure studied in $f(Q)$ gravity \cite{jib} are different from ours because they considered the covariant formulation of $f(Q)$ gravity, obtained in terms of the underlying Levi-Civita connection. Also, in literature, one can find different approaches to study the effective theory, as we can see for instance in \cite{emma, beng}.

The authors of the present comment claimed that they found an extra term $\frac{6 H^2 f_Q}{2\kappa}$ in their effective density and effective pressure. They also claimed that this extra term would offer a correction to the set of energy conditions of $f(Q)$ gravity. However, as we have been shown in this reply, the extra term, in fact, came from their particular way of writing the field equations presented in \eqref{eq01}. Once we prove that their effective density and effective pressure yield to the same set of Friedmann Equations presented by us in \cite{sanjay},  we can repeat the methodology adopted in our article to define $\tilde{\rho}$ and $\tilde{p}$, and compute the energy constraints for $f(Q)$ gravity. Consequently, we respectfully do not expect that any new result would appear in the energy conditions previously found by us in \cite{sanjay}. And, we stand by our approach to derive energy conditions in $f(Q)$ gravity.

%


\end{document}